\documentclass[
 reprint,
superscriptaddress,
amsmath,
amssymb,
aps,
pra,
]{revtex4-1}
\usepackage{comment}
\usepackage{graphicx}% Include figure files
\usepackage{dcolumn}% Align table columns on decimal point
\usepackage{bm}% bold math
\usepackage{color}
\usepackage{multirow}
\usepackage{geometry}
\usepackage{multirow,array}
\usepackage{xfrac}  

\newcolumntype{P}[1]{>{\centering\arraybackslash}p{#1}}
\usepackage{hyperref}
\hypersetup{
     colorlinks   = true,
     citecolor    = magenta,
     linkcolor    = blue,
}
\begin{document}

%\preprint{APS/123-QED}

\title{Phase Properties of Interacting Bosons in Presence of Quasiperiodic and Random Disorder}% Force line breaks with \\

\author{Sk Noor Nabi}
\affiliation{Department of Physics, Indian Institute of Technology Kharagpur, Kharagpur-721302, West Bengal, India}
\email{sknoornabi@phy.iitkgp.ac.in}
\author{Shilpi Roy}
\affiliation{Department of Physics, Indian Institute of Technology Guwahati, Guwahati-781039, Assam, India}
\email[]{roy176121109@iitg.ac.in}
\author{Saurabh Basu}
\affiliation{Department of Physics, Indian Institute of Technology Guwahati, Guwahati-781039, Assam, India}
%\author{Tapan Mishra, Saurabh Basu}%
% %\email{}
%\affiliation{%

\date{\today}% It is always \today, today,

\begin{abstract}
Motivated by two different types of disorder that occur in quantum systems with ubiquity, namely, the random and the quasiperiodic (QP) disorder, we have performed a systematic comparison of the emerging phase properties corresponding to these two cases for a system of interacting bosons in a two dimensional square lattice. Such a comparison is imperative as a random disorder at each lattice is completely uncorrelated, while a quasiperiodic disorder is deterministic in nature. Using a site decoupled mean-field approximation followed by a percolation analysis on a Bose-Hubbard model, several different phases are realized, such as the familiar Bose-glass (BG), Mott insulator (MI), superfluid (SF) phases, and, additionally, we observe a mixed phase, specific to the QP disorder, which we call as a QM phase. Incidentally, the QP disorder stabilizes the BG phase more efficiently than the case of random disorder. Further, we have employed a finite-size scaling analysis to characterize various phase transitions via computing the critical transition points and the corresponding critical exponents. The results show that for both types of disorder, the transition from the BG phase to the SF phase belongs to the same universality class. However, the QM to the SF transition for the QP disorder comprises of different critical exponents, thereby hinting at the involvement of a different universality class therein. The critical exponents that depict all the various phase transitions occurring as a function of the disorder strength are found to be in good agreement with the quantum Monte-Carlo results available in the literature.

\end{abstract}

%We have performed a systematic comparison between a random (RP) and quasi-periodic (QP) potentials on the quantum behaviour of interacting ultracold atoms in optical lattices using the mean-filed approximation (MFA). The site inhomogeneity in the MFA is taken care by defining a new quantity, called indicator for characterizing different quantum phases, such as the Bose glass (BG) and quasi-periodic induced mixed (QM) along with the superfluid (SF) and Mott insulator (MI) phases. Further, a statistical based percolation theory is employed to study the SF cluster formation at the onset of the BG or QM-SF phase transition which shows a maximum value close to the percolation threshold. Finally, the critical properties are investigated using finite size scaling (FSS) analysis and the critical exponents are determined corresponding to various phase transitions for both types of potentials. The FSS analysis reveal that the BG-SF and QM-SF phase transitions have different critical exponents which are found to be in good agreement with the QMC results but almost same fractal behaviour. However, no explicit finite size effect is observed for large system size and the QM and BG phase are boosted in case of QP potential.
\maketitle

%\tableofcontents

\section{\label{sec:level1} Introduction}
The ubiquitous presence of disorder in a plethora of systems demands a thorough investigation of its footprints on the phase properties. Anderson's seminal work on disorder in quantum systems states that a non-interacting system gets localized in presence of a random disorder, resulting in a localization transition from a conducting to an insulating phase \cite{PhysRev.109.1492}. This phenomenon is known as Anderson localization, and is exhibited in larger than two dimensions \cite{PhysRevLett.42.673}. Quite interestingly, such a localization transition can be realized even in one-dimension (1D) in presence of a quasiperiodic (QP) potential, and thus has gained much attention in recent times \cite{roati2008anderson}. The QP disorder belongs to an intermediate regime of a periodic and a fully random potential \cite{SOKOLOFF1985189}. It is characterized by a long-range order, with no translational symmetry \cite{PhysRevLett.53.1951}. In addition to the localization transition in a 1D system, the scenario hosts interesting critical properties, multifractal behaviour at and away from the transition point, critical nature of the eigenspectra \cite{PhysRevLett.51.1198,Siebesma_1987,yao2019critical,deng2019one,PhysRevLett.50.1870,szabo2018non,PhysRevB.50.11365,PhysRevB.96.085119,10.21468/SciPostPhys.4.5.025,PhysRevB.34.2041,PhysRevB.35.1020} etc. 
The presence of the QP disorder in a nearest-neighbour tight-binding Hamiltonian, which is denoted as the Aubry-Andr\'e (AA) model has a fascinating property, namely the self-duality \cite{aubry1980analyticity}. Self-duality yields an intuitive way to understand the (sharp) localization transition. Thus the model does not exhibit a mobility edge, and hence localization transition is energy-independent \cite{Mott_1987}. However, QP disorder goes beyond the realm of the AA model, and shows fascinating phenomena in a variety of systems \cite{an2018engineering,yao2019critical, bodyfelt2014flatbands,wang2020one,PhysRevB.41.5544,PhysRevA.80.021603,PhysRevLett.104.070601,PhysRevB.83.075105,PhysRevLett.114.146601,PhysRevLett.126.040603,PhysRevB.103.184203,deng2019one, PhysRevLett.126.106803}. In particular, QP disorder displays significantly more complex and intriguing results in higher-dimensional systems as well \cite{PhysRevLett.116.140401,PhysRevB.99.054211,PhysRevLett.122.110404, PhysRevB.101.014205,devakul2017anderson}.

In experimental situations, a real material constitutes of a large number of particles, as well as different types of interactions, which lead to various crucial properties. Cold atomic systems have successfully emerged as quantum simulators for realizing many of these complex properties and phase transitions induced by interparticle interactions.
The experimental success in cooling and trapping of cold alkali atoms in optical lattices, where the interplay of such interactions and external fields exhibit novel quantum phenomena, is otherwise inaccessible in conventional condensed matter systems \cite{RevModPhys.82.1225,RevModPhys.70.707}. 

A paradigmatic model which depicts the phases of such an ultracold gas of bosons in an optical lattices is the Bose-Hubbard model (BHM) \cite{PhysRevLett.81.3108}. By tuning the system parameters, a transition from a superfluid phase (SF) to a Mott insulator (MI) can be experimentally observed in a clean environment \cite{Greiner2002}.
% Moreover, the use of an optical dipole trap (ODT) instead of the MOT has provided an edge in retaining the hyperfine degrees of freedom, and thus makes an ultracold atomic system as a spinor gas. As a result of spins being included, a plethora of quantum phases, namely, spin singlet (nematic) MI, polar SF and broken axis symmetry (BA) phases emerge depending upon the sign of the spin dependent on-site interaction potential \cite{PhysRevLett.81.742,ohmi,PhysRevLett.94.110403,PhysRevB.77.014503}.   
Apart from the role of interaction above in stabilizing different phases, the interplay of disorder and interaction assumes a crucial role in hosting the new quantum phases in these systems. Although, an optical lattice is free from impurities, however it is possible to engineer disorder in  optical lattices via several techniques, such as, using speckle laser beams, multicomponent Bose gas or noncommensurate multichromatic lattices etc \cite{PhysRevLett.102.055301,Meldgin2016,PhysRevLett.95.070401,Pasienski2010,PhysRevLett.96.180403}. The presence of disorder in a system of ultracold atoms in optical lattices generates an additional phase, known as the Bose glass (BG) phase which is characterized by finite compressibility and vanishing SF order parameter \cite{PhysRevLett.102.055301,Meldgin2016,PhysRevLett.102.055301,PhysRevLett.98.130404}. In an attempt to ascertain the role of disorder, Fallani {\it{et al.}} created a disordered optical lattice by superimposing an auxiliary laser beam with the main optical lattice potential \cite{PhysRevLett.98.130404}. They have experimentally observed the signature of the BG phase by studying the excitation spectrum of the system as a function of the strength of disorder.

It is worthwhile to mention the seminal work by Fisher {\it{et al.}} in the context of disorder in BHM, and in particular for a diagonal disorder, the BG phase interrupts a direct MI-SF phase transition, although such a direct transition not fundamentally impossible for sufficiently weak disorder \cite{PhysRevB.40.546}. However, evidence suggests that a direct transition from the MI to the SF phase is ruled out in presence of random disorder, and is always intervened by a BG phase \cite{PhysRevLett.103.140402}. 
%Subsequently, extensive efforts have been given to understand the nature and the appearance of the BG phase from a theoretical perspective in the following years. Recently, Pollet {\it{et al.}} gave a strong argument based on a mathematical descriptions and postulated a theorem, known as, the "theorem of inclusions" which identifies the the BG phase as a Griffiths phase dominated by rare-regions \cite{PhysRevLett.103.140402}. They have confirmed the existence of the BG phase upon the destruction of the MI phase, and thereby rule out a direct MI-SF phase transition.
Apart from such fundamental realizations, various numerical techniques, such as, quantum Monte Carlo (QMC) \cite{PhysRevB.98.184206,PhysRevB.84.094507,PhysRevA.98.023628,PhysRevLett.99.050403,PhysRevLett.114.105303,PhysRevLett.87.247006}, stochastic mean-field theory \cite{PhysRevA.81.063643,Bissbort_2009}, Green’s function approach and DMRG \cite{Rapsch_1999,Gerster_2016} etc have been developed to study the BG phase in the disordered cold atomic gases. Moreover, a site dependent mean field approximation (MFA) employs finding of the SF percolating cluster using a percolation analysis to capture the BG phase. The resultant phase diagram appears to be quite similar to that of the QMC results \cite{PhysRevA.99.053610,PhysRevB.85.020501,PhysRevA.91.043632,Niederle_2013}. 

Although a large number of studies so far have focussed on the effect of random disorder present in the BHM, unfortunately it remains highly unexplored how a quasiperiodic potential will influence the quantum phases of interacting bosons in optical lattices. Very recently, the ground state phase diagram of a two-dimensional ultracold Bose gas in optical lattices in presence of QP disorder has been studied using QMC \cite{PhysRevA.91.031604} and MFA \cite{Johnstone_2021,johnstone2022barriers} techniques. In addition to the BG phase, they have found signatures of an additional phase, namely, the quasiperiodic induced mixed (QM) phase along with the BG phase. Besides, there is no signature of a Mott-glass like behaviour present in the system \cite{PhysRevA.91.031604}. In addition, the QP potential in the non-interacting system in two dimensions (2D) also demonstrates the appearance of a mixed spectrum at strong QP disorder strengths \cite{PhysRevB.101.014205}.  

In a general sense, systematic studies of phase transitions are done via critical state analysis. Such an exercise will facilitate a direct comparison between both kinds of disorder that are under the lens in this work.
The critical phenomena in the vicinity of the transition points are characterized by the scaling hypothesis and the universality classes. Moreover, the scaling hypothesis entails data collapse of the curves that denote the critical behaviour corresponding to different system sizes onto a single one via a unique scaling function. This method aids in obtaining the critical exponents corresponding to the transitions.
It can happen that two different critical points may refer to the same universality class if the critical exponents and the scaling forms are identical for both of them \cite{RevModPhys.71.S358, roy2021critical}. Therefore, an in-depth study is necessary to compute the critical points and the exponents corresponding to the observables.

Motivated by such exciting prospects, we shall consider a system of interacting ultracold atoms in a 2D optical lattice in presence of a random and a quasiperiodic (QP) disorder. The main objective is to compare and contrast these two different types of disorder with regard to their effects on the phase properties, and the transitions therein.
Here, we shall use the percolation based MFA to characterize different quantum phases, such as, the QM, BG, MI and the SF phases. Our primary focus will be to study the nature of the QM-SF or BG-SF phase transitions, and find out the critical exponents using a finite-size scaling analysis corresponding to both the random and the QP disorder cases. In section II, we briefly outline our theoretical model, which is followed by the results in section III. Section IV concludes with a brief mention of our key results.

\section{Model and approach}\label{model}
Here we consider a BHM that describes the general properties of ultracold atoms loaded in a two dimensional (2D) square lattice in the presence of the random and quasiperiodic potentials. The corresponding Hamiltonian is written as \cite{PhysRevLett.81.3108,PhysRevLett.114.105303,Niederle_2013},
\begin{equation}
H=-t\sum\limits_{<i,j>}(\hat{a}^{\dagger}_{i}\hat{a}_{i}+
h.c.)-\sum\limits_{i}(\mu-\epsilon_{i})\hat{n}_{i}
+ \sum\limits_{i}\frac{U}{2}\hat{n}_{i}(\hat{n}_{i}-1)
\label{Ham}
\end{equation}
where $\hat{n}_{i}$ represents the occupation number operator, that yields the number of bosons at a lattice site $i$. The number of bosons increases (decreases) via $\hat{a}^{\dagger}_{i}$ $(\hat{a}_{i})$, which denotes the creation (annihilation) operator at the lattice site $i$. The hopping amplitude between the nearest-neighbor lattice sites is denoted by $t$. $\langle ij \rangle$ represents a pair of nearest-neighbour lattice sites $i$ and $j$. The particle density can be controlled via the chemical potential $\mu$, while the repulsive interaction strength between particle densities at a particular site is denoted by $U$. 
$\epsilon_{i}$ represents the on-site energy introduced at the lattice site $i$. In this work, we shall consider two different forms of $\epsilon_{i}$ that depict the presence of such random  and  quasiperiodic (QP) disorder. 
For the random disorder, we choose a uniformly distributed energies, $\epsilon_{i}$ from a box distribution [-$\Delta$, $\Delta$], where $\Delta$ denotes the strength of the random on-site potential \cite{PhysRevLett.114.105303,Niederle_2013}. 
For the other choice, $\epsilon_{i}$ represents the two dimensional QP disorder at a site in a 2D square lattice, $i\in (n,m)$ which is given by, \cite{PhysRevA.91.031604,Johnstone_2021,johnstone2022barriers,Deng2009}
\begin{equation}
\epsilon_{i}=-\lambda\bigg( \cos \big[ 2 \pi \beta (n_{i}+m_{i})+\alpha \big]+ \cos \big[2 \pi \beta (n_{i}-m_{i})+\alpha\big] \bigg)
\end{equation}
where $\lambda$ denotes the strength of the potential, $n,m$ being the site indices, and $\alpha=[0,2\pi]$ is a phase factor. $\beta$ determines the periodicity of the quasiperiodic potential. In our work, we have assumed
$\beta=M/L$ where $M$ and $L$ are co-prime integers. Our assumption is based on the continued fraction expansion \cite{Lang} which allows one to incorporate the periodic boundary conditions on a square lattice, $L \times L$. We choose $\beta=\frac{1}{\sqrt{2}}$, which fixes the possible system sizes to be $L=55, 99$ and $239$ etc \cite{note1}. In the following, we often denotes random disorder by $\Delta$ and the QP disorder by $\lambda$.

In order to study the phase transition between different quantum phases, we shall employ the mean field approximation (MFA) to decouple the hopping term as follows \cite{PhysRevB.85.214524,PhysRevA.63.053601},
\begin{eqnarray}
\hat{a}^{\dagger}_{i}\hat{a}_{j} \simeq \langle
\hat{a}^{\dagger}_{i} \rangle \hat{a}_{j}
+\hat{a}^{\dagger}_{i}\langle
\hat{a}_{j}\rangle-\langle
\hat{a}^{\dagger}_{i}\rangle \langle
\hat{a}_{j}\rangle
\label{deco}
\end{eqnarray}
where $\langle\hspace{2mm} \rangle$ denotes the equilibrium value of an
operator and defining the SF order parameter at site $i$ as, $\psi_{i}= \langle \hat{a}_{i}\rangle$.
Now substituting the superfluid order parameter in Eq.(\ref{Ham}), the BHM can be written as a sum of single site Hamiltonians as,
\begin{equation}
H=\sum\nolimits_{i}H^{MF}_{i} 
\end{equation}
where,
\begin{eqnarray}
H^{MF}_{i}&=&-zt(\phi^{*}_{i}\hat{a}_{i}+h.c.)+zt\phi^{*}_{i}\psi_{i}-(\mu-\epsilon_{i})\hat{n}_{i}\nonumber\\ &+&
\frac{U}{2}\hat{n}_{i}(\hat{n}_{i}-1).
\label{mf}
\end{eqnarray}
Here $\phi_{i}=(\sfrac{1}{z})\sum\nolimits_{j}\psi_{j}$ and the sum over $j$ includes all nearest-neighbors of a site $i$ in a square lattice, $z$ is the coordination number, $z=2d=4$, $d$ being the lattice dimension.

Randomly chosen initial local order parameters are used to get local ground state by diagonalizing the mean field Hamiltonian [Eq.(\ref{mf})]. This process is repeated at all lattice sites, and hence the state corresponding to the entire lattice generates a global ground state. The self-consistency is checked at each step of the process, until the global order parameters  converge within a certain accuracy (in our computation, the accuracy is taken an $0.001$). By using the global ground state $|\Psi_{G}\rangle$ of the Hamiltonian, various physical quantities, such as, the superfluid order parameter $\psi_{i}$, the occupation number, $\rho_{i}$, and the compressibility $k_{i}$ are computed using the following relations, 
\begin{eqnarray}
\psi_{i} = \langle \Psi_{G}|a_{i}|\Psi_{G}\rangle \\
\rho_{i} = \langle \Psi_{G}|n_{i}|\Psi_{G}\rangle\\
\bar{\kappa} =\Big[ \sum_{i=1}^{L^{2}}[\rho_{i}^{2}-(\rho_{i})^{2}\Big]
\end{eqnarray}
In addition to that, the average SF order parameter ($\bar{\Psi}$) and the compressibility ($\bar{\kappa}$) are defined via,
\begin{eqnarray}
%\begin{equation}
\bar{\Psi} =\Big[\bigg(\frac{1}{L^{2}}\bigg) \sum_{i=1}^{L^{2}} \psi_{i}\Big]_{sample} \\
\bar{\kappa} =\Big[\bigg(\frac{1}{L^{2}}\bigg) \sum_{i=1}^{L^{2}}[\rho_{i}^{2}-(\rho_{i})^{2}]\Big]_{sample}
\end{eqnarray}
where the $sample$ in the subscript refers to the fact that results are averaged over different disorder realizations. For a random disorder, the notion of $sample$ denotes various random configurations of the onsite disorder in the range $[-\Delta: \Delta]$, while for QP disorder, several values of $\alpha$ contained within $[0:2\pi]$ have been considered by us.
For our calculations, we have taken $sample=100$ and $sample=50$ different realizations for the random and QP disorder respectively. However, other values for $sample$ have also been used on certain occasions. To facilitate a comparison between the two types of disorder, we have used three different representative disorder strengths, namely, $\lambda/U$ and $\Delta/U$ to be $= 0.18,~0.30,$ and $0.55$ which, respectively denote weak, moderate, and high disorder strengths. 
\begin{figure}[h!]
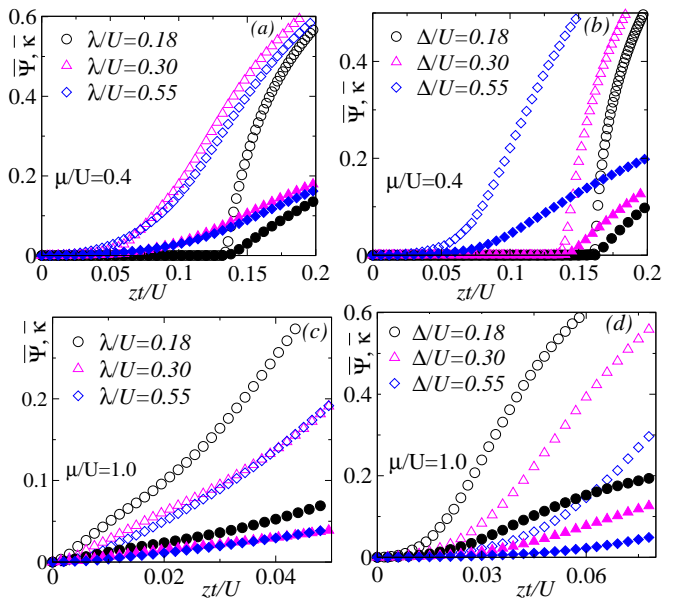

\centerline{\hfill
\includegraphics[width=0.247\textwidth]{Fig1a.eps}
\hfill
\hfill
\includegraphics[width=0.247\textwidth]{Fig1b.eps}
\hfill
}
\centerline{
\hfill
\includegraphics[width=0.24\textwidth]{Fig1c.eps}
\hfill
\hfill
\includegraphics[width=0.24\textwidth]{Fig1d.eps}
\hfill
}
\caption{The variation of the average SF order parameter $\bar{\Psi}$ (open symbols) and compressibility $\bar{\kappa}$ (filled symbols) for the MI-BG-SF ($\mu/U=0.4$)  and QM-SF ($\mu/U=1.0$) phases are shown corresponding to QP disorder in (a)-(c) and random disorder in (b)-(d).}
\label{fig:sf_comp}
\end{figure}

\section{Results}\label{res}

In the absence of an onsite energy, the model Hamiltonian (\ref{Ham}) coincides with the homogeneous BHM, and its underlying physics is well understood. Due to the competition between the hopping strength ($t$) and the repulsive interaction ($U$), the BHM undergoes a phase transition from the MI to the SF phase. However, in presence of any sort of inhomogeneity, the system experiences the intrusion of additional phases in between the SF and MI phases, namely, the BG or the QM phases. 
%To explore the effects of both QP and RP disorder, and importantly a comparison between them we shall start our discussion with the behaviour of the averaged order parameter and the compressibility and then turn our attention to the percolation scenario. Our primary aim is to investigate the BG or the QM phases, for which 
To this effect, we shall consider two representative values of the chemical potential, such as $\mu/U=0.4$ specifically while focussing on the BG phase, and $\mu/U=1.0$ for discussing the QM phase. 
\subsection{SF order parameter and compressibility}
%In a disorder free scenario, there is a clear boundary separating the MI from the SF phase and thus a sharp phase can be observed. In a phase diagram defined by the $\mu/U$-$zt/U$ plane, the system shows formation of  Mott lobes till a certain hopping amplitude. However, by increasing the hopping strength, the superfluid behaviour sets in. This point yields the critical hopping strength corresponding to the phase transition. The inclusion of onsite energies induces inhomogeneities whence the lattice becomes energy-dependent and unstable. In order to gain a deeper insight, we need to focus on the phases that appear in the presence of inhomogeneities. 

In general, the MI, BG, QM and the SF phases can be distinguished by looking into the behaviour of the order parameters. It is observed that the MI phase can be identified with vanishing of both the SF order parameters and the compressibility. Thus each lattice site hosts the same (and integer) number of bosons for the MI phase. In contrast, the SF phase comprises of non-integer occupation densities, finite superfluid order parameter, and non-vanishing values for the compressibility. In the presence of the QP and the random disorder, additional phases, such as, the BG and the QM phases appear, and demonstrate richer phase properties. A non-integer occupation number emerges corresponding to the BG phase; however, the superfluid order parameter remains zero. On the other hand, the QM phase looks more insulating in nature, and less like the SF phase. Thus it behaves as a pseudo-SF or weak-SF phase. 

In order to understand the nature of the BG or the QM phases in presence of both types of potential, we calculate the average SF order parameter and the compressibility corresponding to different QP and random strengths, as shown in the  Fig.~\ref{fig:sf_comp} (top row) for BG and Fig.~\ref{fig:sf_comp} (bottom row) for QM phases.

For a particular value of $\mu$, namely, $\mu/U=0.4$, the variation of $\bar{\Psi}$ and $\bar{\kappa}$ corresponding to the QP disorder in Fig.\ref{fig:sf_comp} (a) and the random disorder in Fig.\ref{fig:sf_comp} (b) follow a similar behaviour as a function of the hopping strength, $t$ (scaled by $U$, namely, $zt/U$). Noticeably, a smaller QP strength, that is, $\lambda/U=0.18$ becomes sufficient to destroy the MI phase and induce the BG phase, resulting in the MI-SF phase transition to occur at a relatively smaller value of the hopping strength, namely, at $zt_{c}/U \approx 0.134$. In contrast, a same value of random disorder strength, that is, $\Delta/U=0.18$ indicates that the MI-SF phase transition occurs at a larger hopping strength, given by at $zt_{c}/U \approx 0.165$. With  gradual increase of the disorder strength, the BG phase encroaches in between the MI and the SF phases by gradually suppressing the Mott insulating phase, thereby lowering the values for the critical transition points.  

While a slightly higher value of QP strength almost destroys the MI phase, leaving the system to only consist of the BG and SF phases at $\lambda/U \ge 0.3$. For the random case, a similar trend is also observed, with a larger value namely, $\Delta/U \ge 0.55$ makes the MI phase unstable, and hence aids the BG phase to encroach into the MI regime. Therefore, one can infer that, corresponding to same strengths of the potentials, the quasiperiodic disorder is more efficient in inducing the BG phase in the system as compared to the case for the random disorder .

For $\mu/U=1.0$, the behavior of $\bar{\Psi}$ and $\bar{\kappa}$ indicate different features corresponding to the QP disorder in Fig.\ref{fig:sf_comp} (c) and random disorder in Fig.\ref{fig:sf_comp} (d). In particular, we obtain the SF order parameter, $\bar{\Psi}\approx 0.2$ and the compressibility, $\bar{\kappa}<0.1$ at a QP disorder value  $\lambda/U=0.18$, while $\bar{\Psi}\approx 0.6$ and $\bar{\kappa}<0.2$ are realized corresponding to the random potential strength, $\Delta/U=0.18$. With increasing $\lambda/U$, $\bar{\Psi}$ and $\bar{\kappa}$ become vanishingly small at $zt_{c}/U \approx 0.04$, thereby pointing towards an insulator like phase. Such a phase seems to be more dominant than the SF phase, and possibly gives an insight on the QM-SF phase transition. Whereas, with increasing $\Delta/U$, the order parameter and the compressibility acquire moderate values, and thus indicate a BG-SF phase transition.  

Although by looking at the average SF order parameter and the compressibility, one can get the qualitative information about  the critical $zt_{c}/U$ separating different quantum phases. However, it fails to provide more information about the appearance of the QM phase, and accurately locate the the critical tunneling strengths for different phase transitions. One possible shortcoming, that may be involved with the site inhomogeneities here, is that all the informations on the MI, BG, QM and the SF phases may partially or completely be lost once the averaging over different configurations is done. Thus we suggest of a more elegant technique below. 

\begin{figure}[t!]
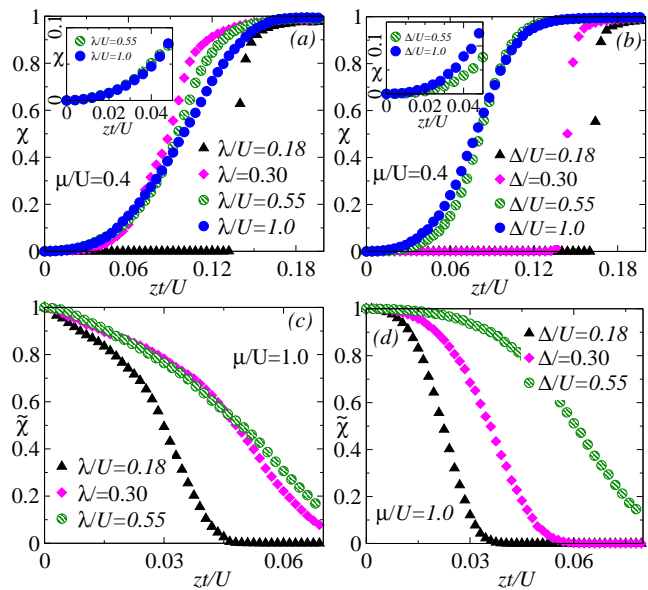

\centerline{\hfill
\includegraphics[width=0.24\textwidth]{Fig2a.eps}
\hfill
\hfill
\includegraphics[width=0.24\textwidth]{Fig2b.eps}
\hfill
}
\centerline{\hfill
\includegraphics[width=0.24\textwidth]{Fig2c.eps}
\hfill
\hfill
\includegraphics[width=0.24\textwidth]{Fig2d.eps}
\hfill
}
\caption{The variation of the indicator ($\chi$) as a function of the hopping strength $zt/U$ for the MI-BG-SF ($\mu/U=0.4$) and QM-SF ($\mu/U=1.0$) phases is shown corresponding to QP disorder in (a)-(c) and random disorder in (b)-(d).}
\label{fig:indicator}
\end{figure}

\subsection{Indicators of MI, BG, QM and SF phases}
General conclusions achieved from $\bar{\Psi}$ and $\bar{\kappa}$ study indicate that it is reasonable to define an observable, namely, an indicator ($\chi$), which is given by \cite{apurba,Nabi_2016}
\begin{equation}
\chi=\frac{{\rm{sites~with}}~ \psi_{i} \neq 0 ~{\rm{and}} ~\rho_{i} \neq integer}{{\rm{total~number~of~sites}}}.
\label{chi}
\end{equation}
In addition, we can also define another quantity, namely $\tilde{\chi}$ by changing the numerator of the above Eq.(\ref{chi}) to the {\it{"\rm{sites with} $\psi_{i}=0~and~\rho_{i}={\rm{integer}}"$}}.  These definitions imply that the MI phase corresponds to $\tilde{\chi}=1$, while the SF phase is characterized by $\chi=1$. Further, they are related via $\tilde{\chi}+\chi=1$. Any intermediate value, such as $0<\chi<1$ signifies the presence of the BG phase. For numerical convergence, we have set the tolerance of $\psi_{i}<O(10^{-1})$ and $\rho_{i}=m \pm \delta$, where $\delta$ is of $O(10^{-3})$ and $m$ is an integer for the MI phase. 
The variations of $\chi$ for the parameters discussed above, that is for $\mu/U=0.4$ are shown in Fig.~\ref{fig:indicator} (a-b), and $\tilde{\chi}$ for $\mu/U_{0}=1.0$ appear in Fig.~\ref{fig:indicator} (c-d) as a function of $zt/U$ corresponding to both types of potentials. 

For $\mu/U=0.4$, it is observed that the MI phase exists up to a value 
$zt/U \approx 0.132$ for $\lambda/U=0.18$ [Fig.~\ref{fig:indicator}(a)],
 and $zt/U \approx 0.162$ for $\Delta/U=0.18$ [Fig.~\ref{fig:indicator}(b)]. Such numbers have been inferred from the plots on averaged SF order parameter and the compressibility as well. Beyond these critical hopping strengths, the values of $\rho_{i}$ turn from integer to fractions, giving rise to the SF phase characterized by finite values of  $\psi_{i}$. With increasing the strengths of the disorder to $\lambda/U=\Delta/U=0.3$, it is seen that, $\chi$ becomes finite for a relatively lower value of $zt/U=0.028$, and $0.134$, respectively for the QP and random potentials. This indicates that the appearance of the BG phase is more pronounced in the QP compared to the random case, which was observed earlier. 

Further increase of the disorder potential, for example to values, such as $\lambda/U=\Delta/U \ge 0.55$, the presence of the BG phase is stabilized over a region with larger values of $zt/U $. Also, the MI phase corresponds to a narrow region of $zt/U \simeq 0.020$ which is shown in Fig.~\ref{fig:indicator} (inset). Thus the MI phase becomes unstable in presence of stronger random disorder, since the order parameters at individual sites delocalize across the lattice, thereby resulting in the presence of the BG and the SF phases only. 

For $\mu/U=1.0$, we have plotted $\tilde{\chi}$ (instead of $\chi$) which will be more useful to witness the signature of the QM phase corresponding to different values of the potential strengths. It shows that $ \tilde{\chi} \neq 0$ till $zt/U \approx 0.046$ for $\lambda/U=0.18$ [Fig.~\ref{fig:indicator}(c)], and $zt/U \approx 0.04$ for $\Delta/U=0.18$ [Fig.~\ref{fig:indicator}(d)] . Further increase upto a value $\lambda/U=\Delta/U \ge 0.3$, results in $\tilde{\chi} \neq 0$ at higher values of $zt/U$, which implies that the insulating phase is now distributed over a large region of the parameter space, thus paving the way for the appearance of the QM phase. The QM phase behaves mostly like an insulating phase, with very weak superfluid character. Thus one can expect that the majority of the lattice sites will host integer occupation densities, except for a tiny region where strong particle correlations will exist. Consequently, we infer that the QM phase corresponds to $\tilde{\chi}>0.5$, which is a patch in the SF region. However, it is too early to talk about the QM phase, instead we need a concrete analysis in order to understand the formation of the SF region in the vicinity of the insulating phase particularly in presence of the QP disorder. 

The above observations suggest that the determination of a precise transition point from the MI-BG phase can be done via $\chi$ or $\tilde{\chi}$. But the presence of the QM phase, and hence the BG or the QM-SF phase boundary remain challenging. We shall turn our attention to the percolation analysis to overcome these hurdles in the next subsection. 

\begin{figure}[h!]
\centerline{\hfill
\includegraphics[width=0.15\textwidth]{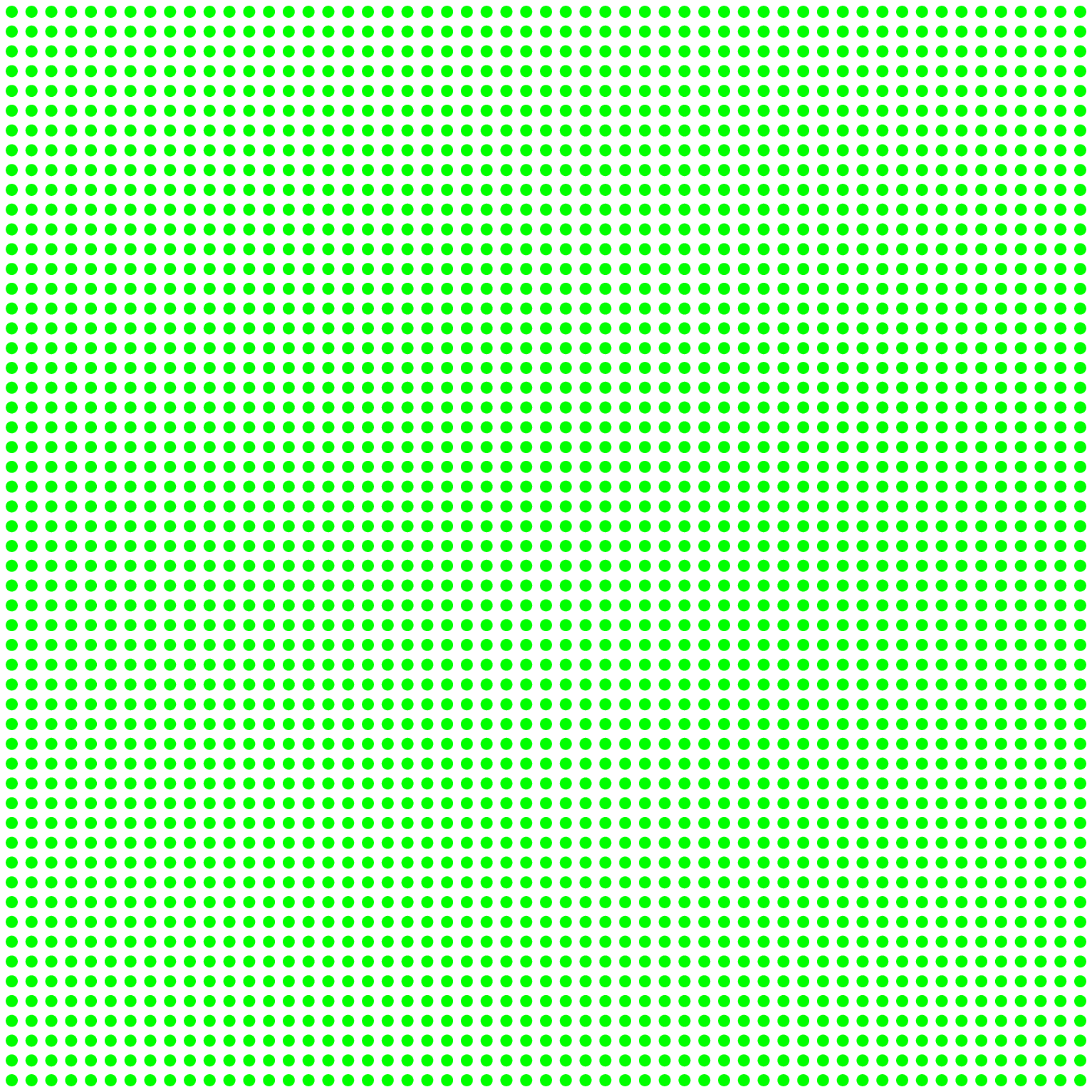}
\hfill
\hfill
\includegraphics[width=0.15\textwidth]{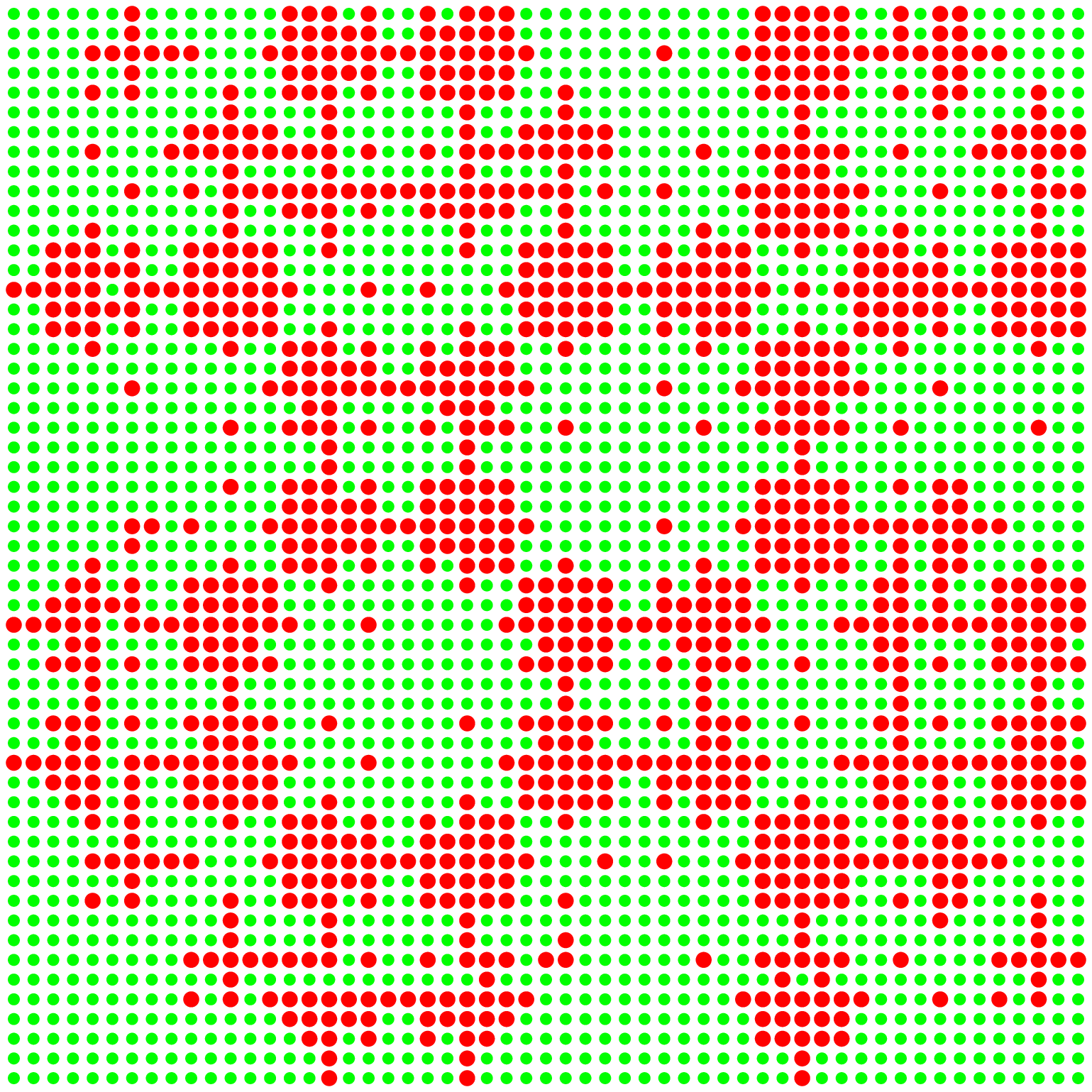}
\hfill
\hfill
\includegraphics[width=0.15\textwidth]{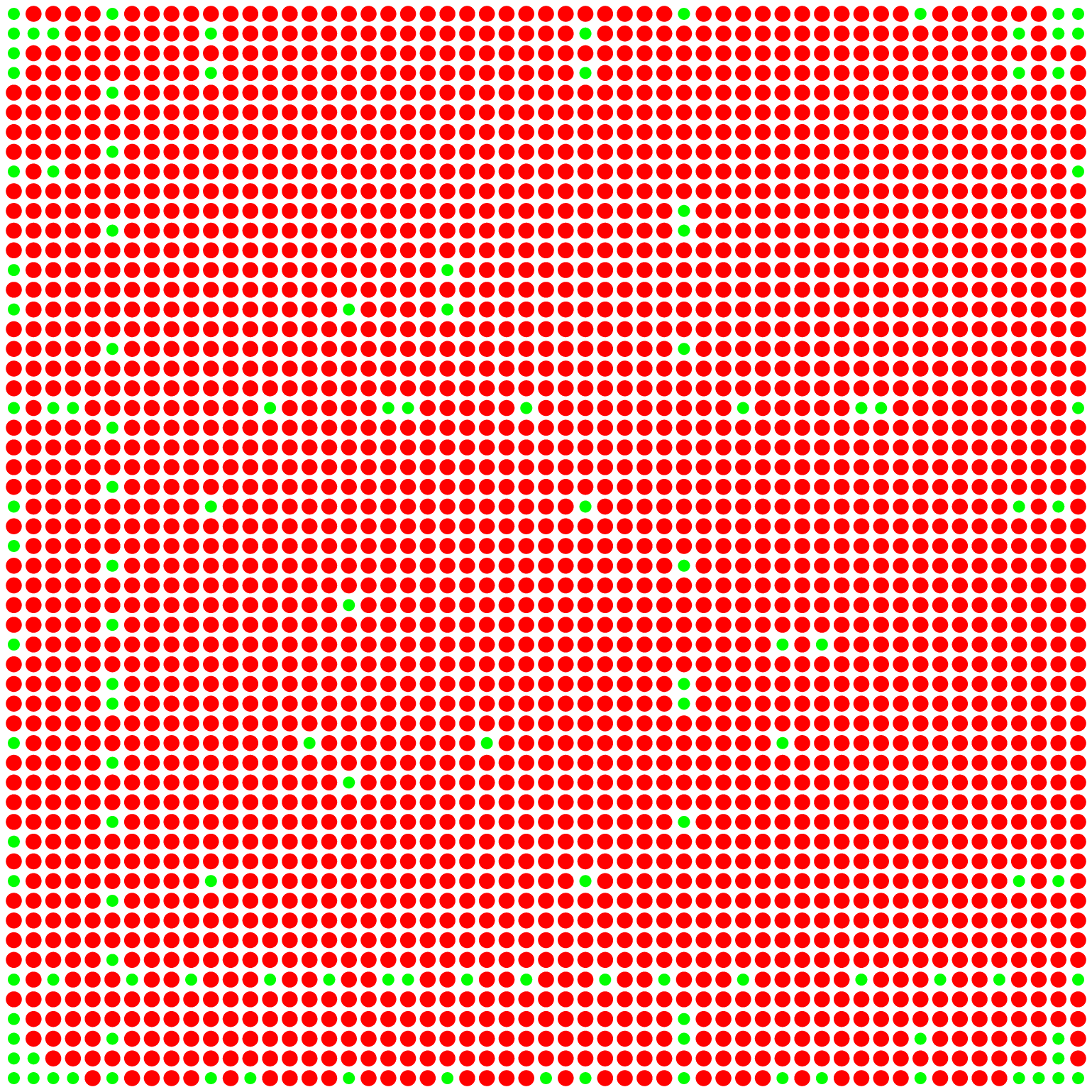}
\hfill
}
\centerline{(a)~ MI \hspace{1.5cm}  (b)~ BG  \hspace{1.0cm} (c)~ SF } 
\centerline{\tiny{$[\tilde{\chi}=1.0, P_{perc}=0.0]$\hspace{0.2cm}$[\chi=0.369, P_{perc}=0.0]$\hspace{0.2cm}  $[\chi=0.958, P_{perc}=1.0]$}}
%%%%%
\centerline{\hfill
\includegraphics[width=0.15\textwidth]{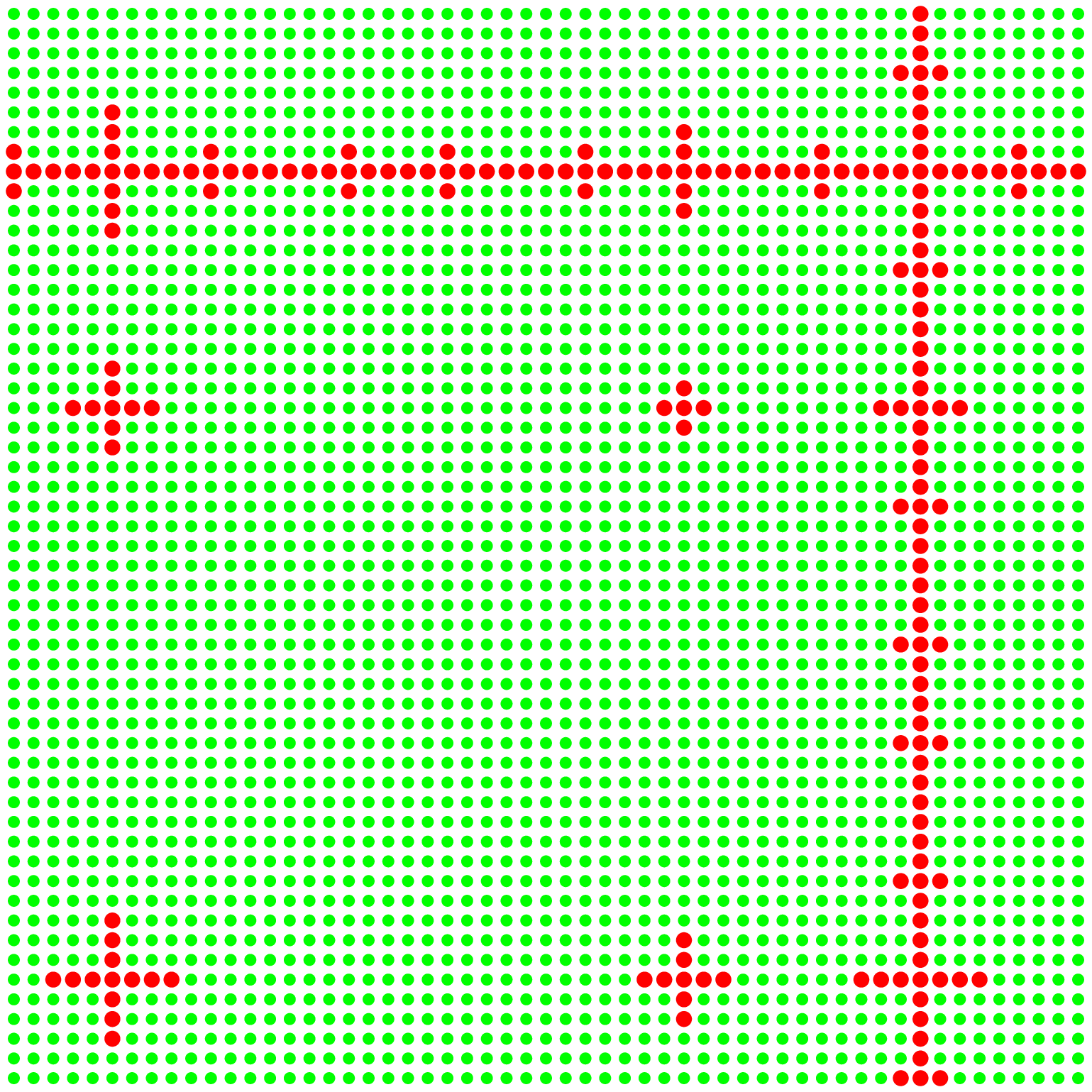}
\hfill
\hfill
\includegraphics[width=0.15\textwidth]{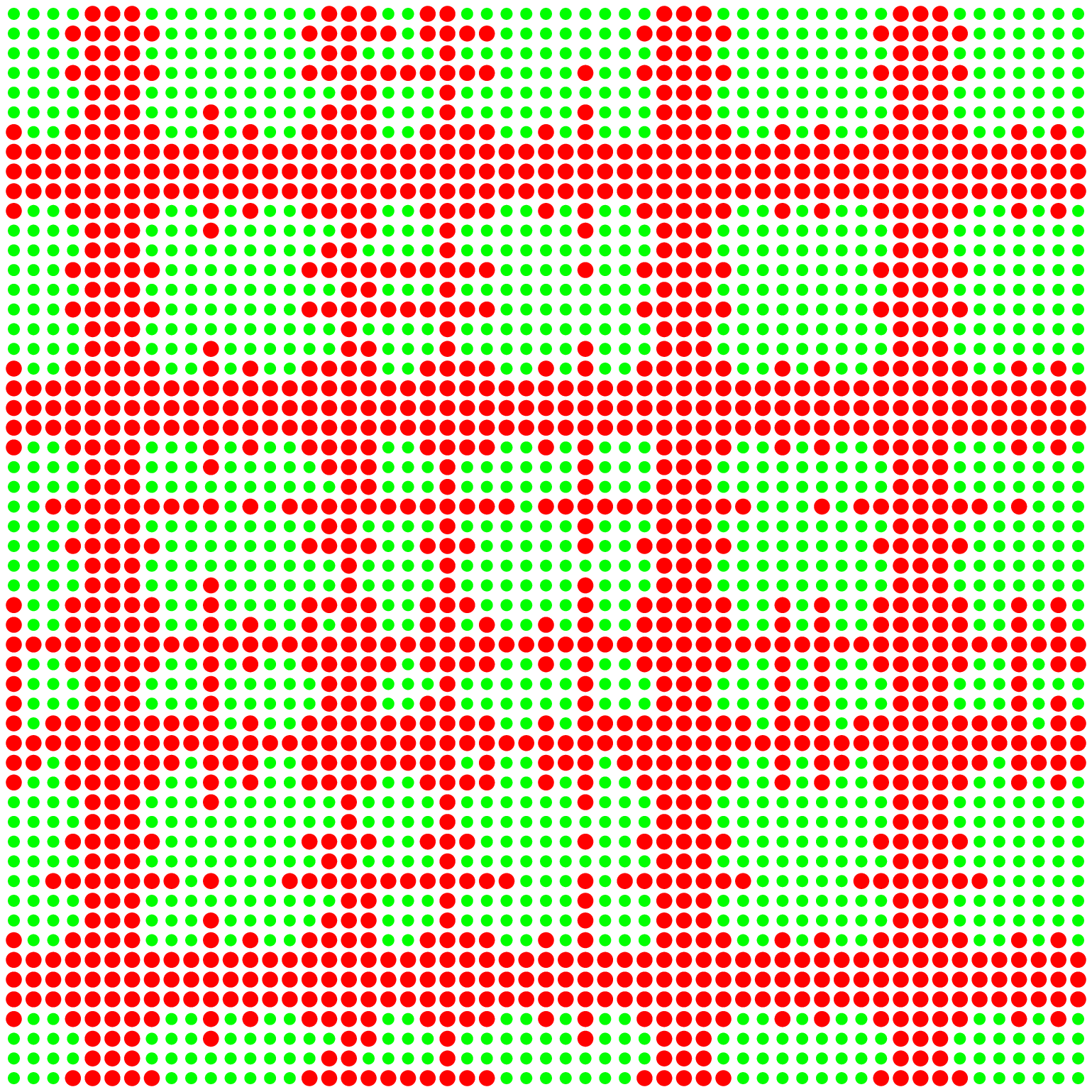}
\hfill
\hfill
\includegraphics[width=0.15\textwidth]{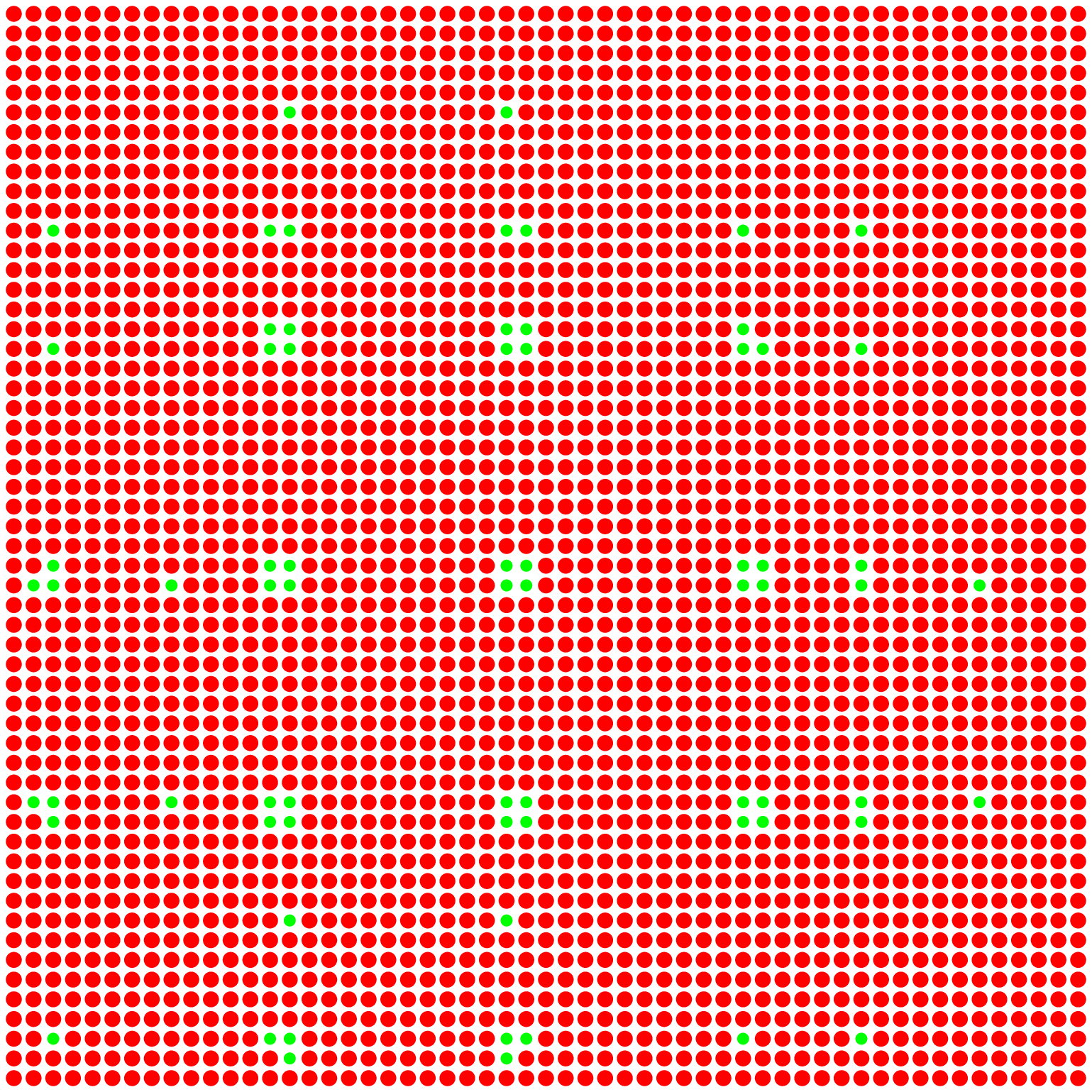}
\hfill
}
\centerline{(d)~ QM \hspace{0.6cm}  (e)~ QM-SF  \hspace{0.7cm} (f)~ SF } 
\centerline{\tiny{$[\tilde{\chi}=0.936, P_{perc}=0.813]$\hspace{0.0cm}$[\tilde{\chi}=0.483, P_{perc}=0.995]$\hspace{0.0cm}  $[\chi=0.976, P_{perc}=1.0]$}}
%%%%%
\centerline{
\hfill
\includegraphics[width=0.15\textwidth]{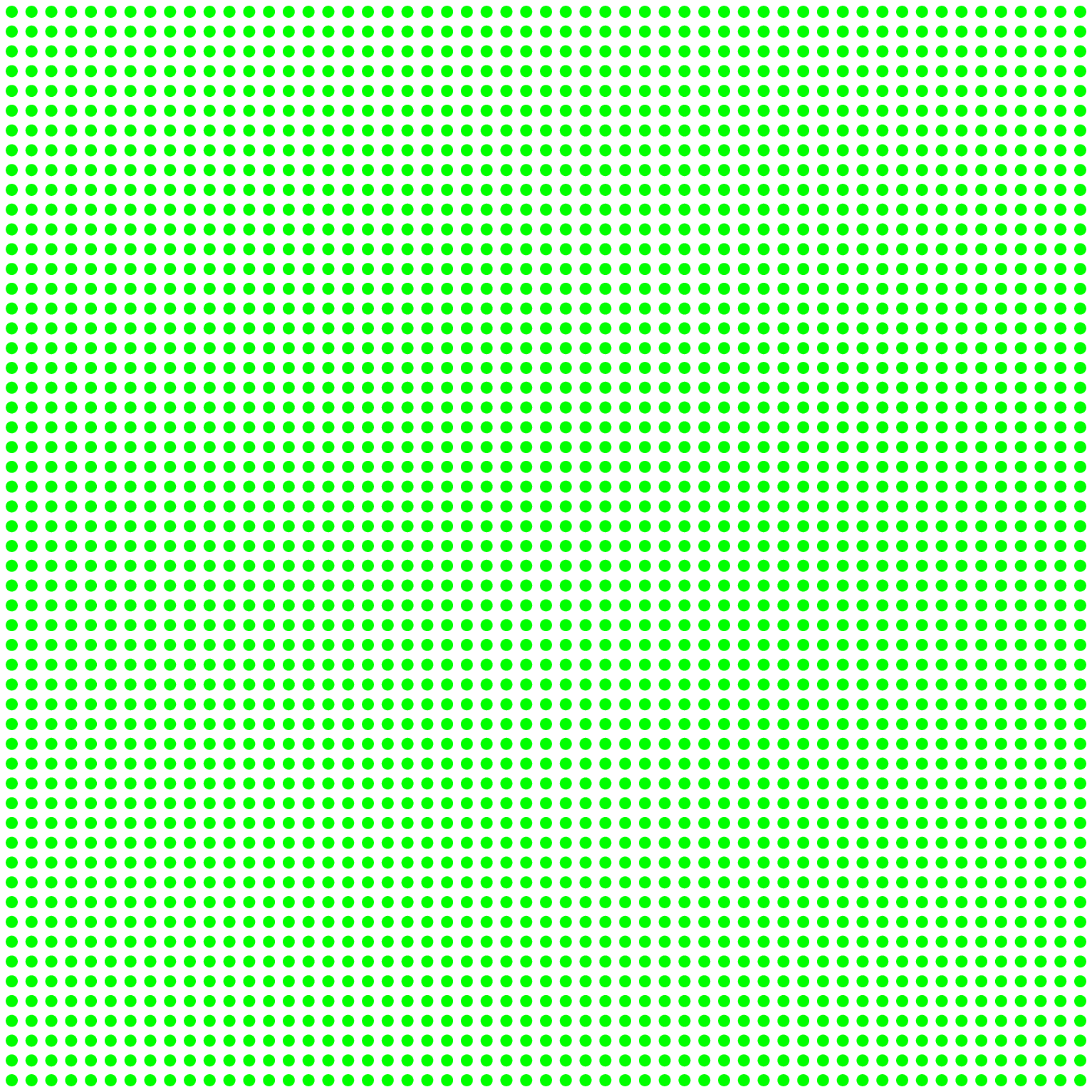}
\hfill
\hfill
\includegraphics[width=0.15\textwidth]{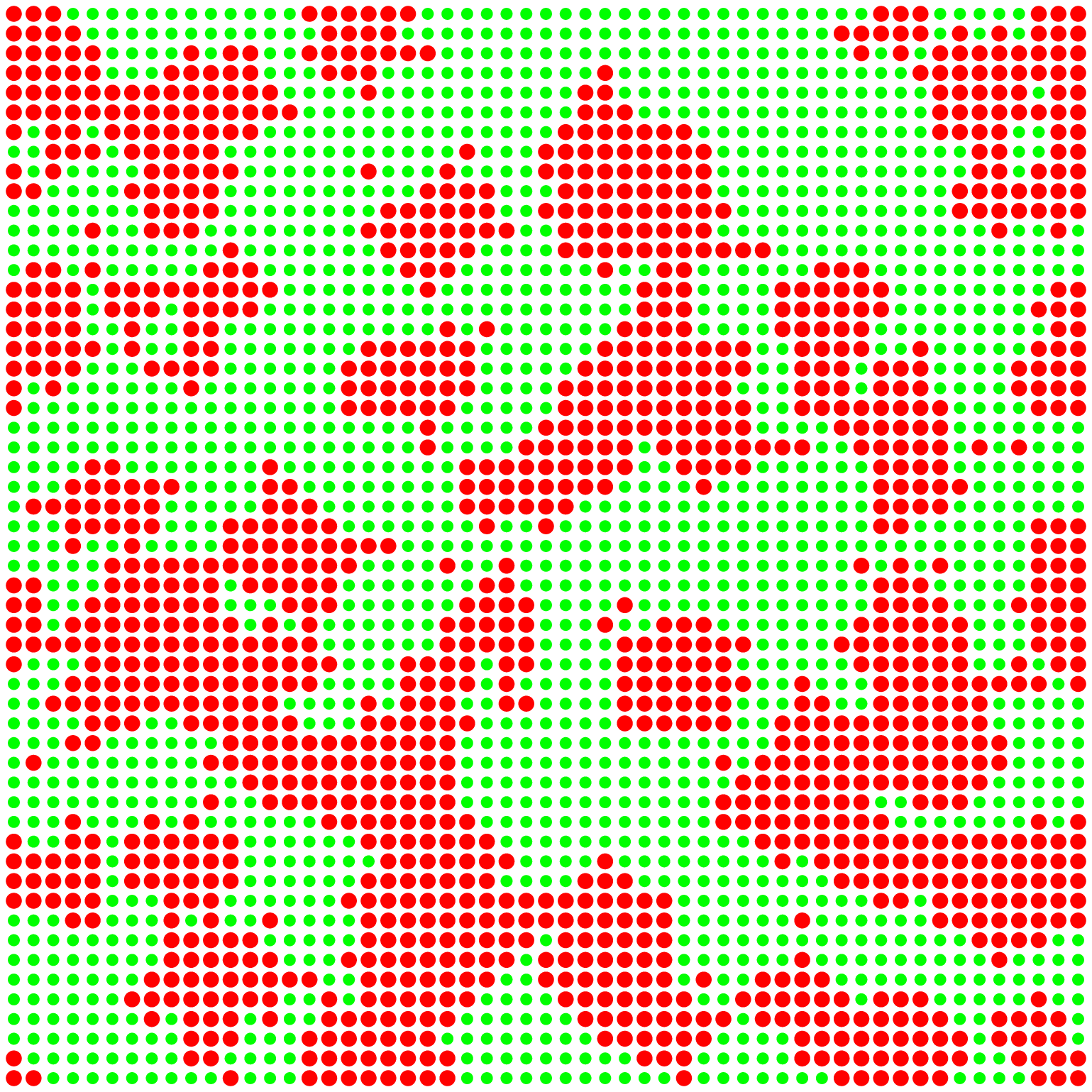}
\hfill
\hfill
\includegraphics[width=0.15\textwidth]{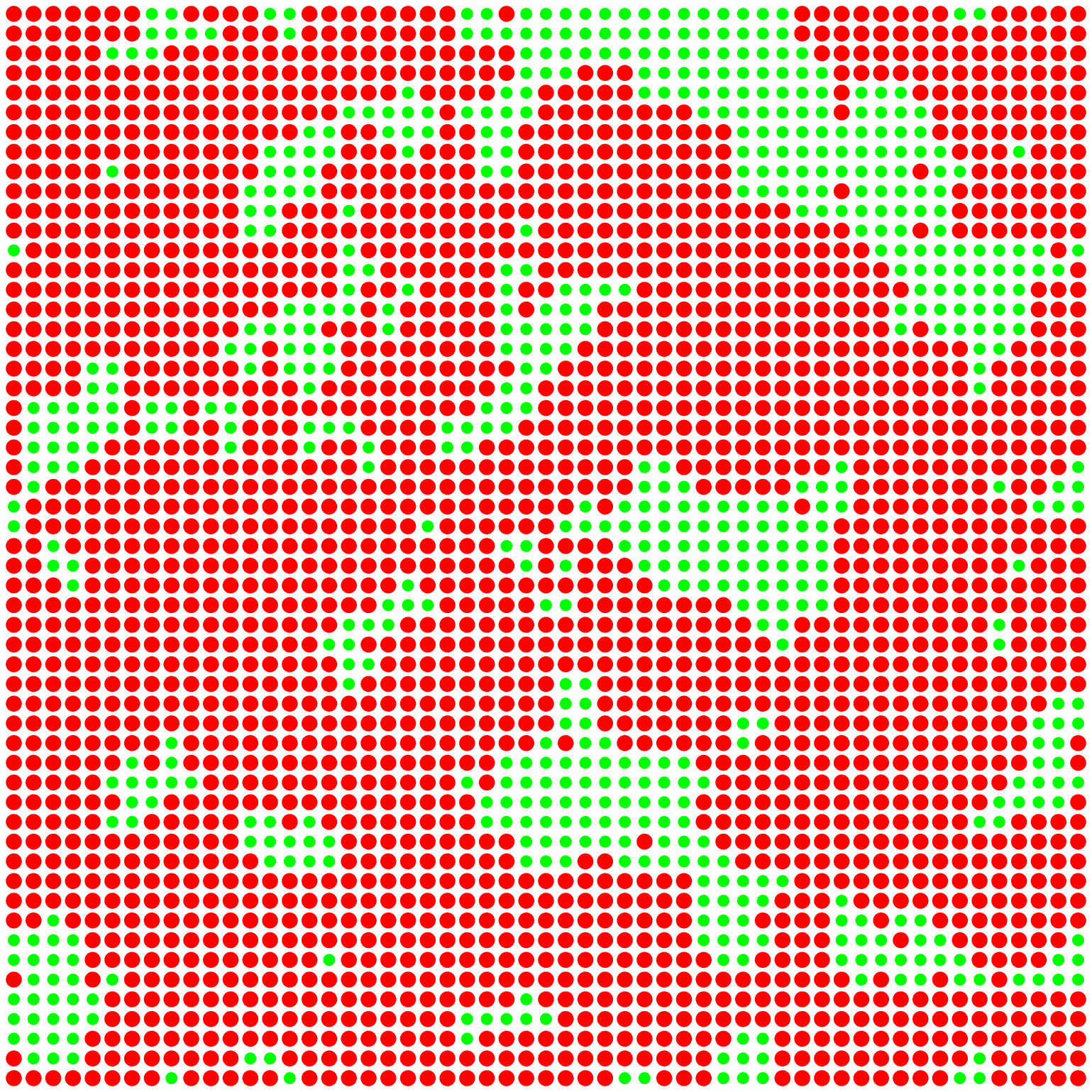}
\hfill
}
\centerline{(g)~MI\hspace{1.5cm}  (h) ~BG \hspace{1cm} (i) ~SF}
\centerline{\tiny{$[\tilde{\chi}=1.0, P_{perc}=0.0]$\hspace{0.2cm}$[\tilde{\chi}=0.534, P_{perc}=0.0]$\hspace{0.2cm}  $[\chi=0.781, P_{perc}=0.946]$}} 
\caption{The real space plots of the occupation densities, $\rho_{i}$
    for lattice size $L\times L=55\times 55$ for the MI, QM, BG and SF phases corresponding to quasiperiodic and random potentials. Top row shows the phase evolution from MI ($zt/U=0.0$) (a) - BG ($zt/U=0.101$) (b) - SF ($zt/U=0.12$) (c) at $\mu/U_{0}=0.4$ for $\lambda/U=0.2$. Middle row shows the phase evolution from QM ($zt/U=0.005$) (d) - Near QM-SF ($zt/U=0.025$) (e) - SF ($zt/U=0.04$) (f) at $\mu/U_{0}=1.0$ for $\lambda/U=0.2$.
Bottom row shows the phase evolution from MI ($zt/U=0.001$) (g) - BG ($zt/U=0.017$) (h) - SF ($zt/U=0.022$) (i) at $\mu/U_{0}=1.0$ for $\Delta/U=0.2$.}
\label{fig:realspace}
\end{figure}

\subsection{Percolation appearance and cluster size distribution}
It is now understandable that the presence of the QP or the random disorder induces two different phases in the system as a function of $zt/U$ for a particular value of $\mu/U$. Since we are interested in the BG and the QM phases, our primary focus will be on the BG-SF and the QM-SF phase transitions, and hence compute the critical points to enumerate the nature of the BG and the QM phases. 
The BG phase constitutes a number of isolated SF clusters that are surrounded by the MI clusters. The SF cluster is an island that is formed by the non-integer occupation densities, while the MI cluster is formed with sites having integer occupation densities. Thus the onset of superfluidity begins when these separated SF clusters coalesce together to percolate or span through the entire lattice for the first time. The QM phase hosts a situation where the majority of the lattice sites belong to the MI cluster, with at least one SF percolating cluster. Therefore, our requirements culminate into finding out the first percolating cluster exhibited in the SF phase corresponding to both the types of disorder. 

For a better understanding, visual representations of
the real space density plots, $\rho_{i}$ are shown in Fig.~\ref{fig:realspace} corresponding to single realization of the QP and the random disorder at a particular value, namely, $\lambda/U=\Delta/U=0.2$. The light green circles indicate $\rho_{i
}=$integer, while the red circles are for $\rho_{i}\neq $integer values. The horizontal (vertical) axis is the lattice site along the $x$ ($y$) direction. In all the three cases, there is a clear evolution from one quantum phase to another. Let us discuss them below.

Fig.~\ref{fig:realspace} (top row) shows the phase evolution from MI ($zt/U=0.0$) (a) - BG ($zt/U=0.1$) (b) - SF ($zt/U=0.12$) (c) at $\mu/U_{0}=0.4$ for $\lambda/U=0.2$. The MI phase in (a) now gradually transforms into the BG phase, having scattered or island-like SF clusters that are surrounded by the MI clusters (b). Finally these SF clusters percolate into the superfluid phase resulting the formation of a spanning cluster across the lattice (c).

\begin{figure}[h!]
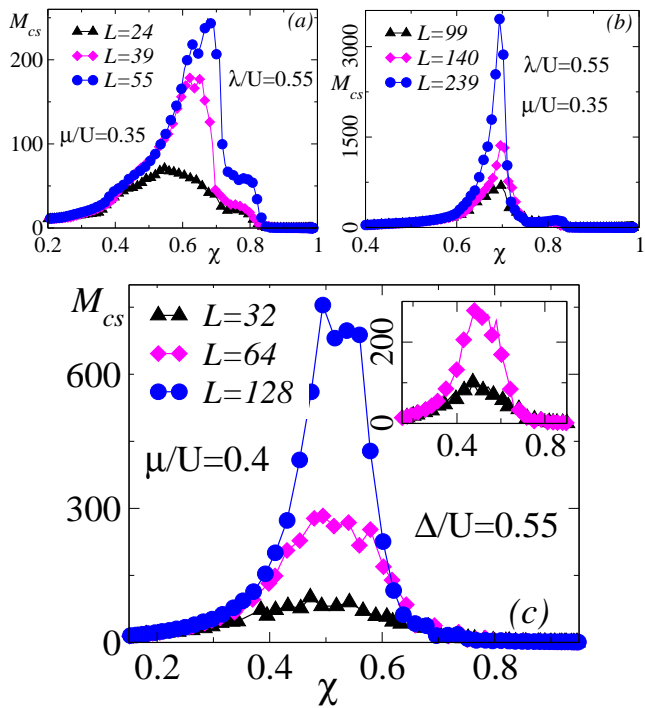

\centerline{\hfill
\includegraphics[width=0.24\textwidth]{Fig4a.eps}
\hfill
\hfill
\includegraphics[width=0.24\textwidth]{Fig4b.eps}
\hfill}
\vspace{2mm}
\centerline{\hfill
\includegraphics[width=0.4\textwidth]{Fig4c.eps}
\hfill
}
\caption{The Mean cluster size, $M_{cs}$ as a function of $\chi$ for different lattice size in the BG-SF phase corresponding to QP disorder in (a)-(b) and random disorder in (c).}
\label{fig:meanclustersizebg}
\end{figure}

Fig.~\ref{fig:realspace} (middle row) displays the phase evolution from QM ($zt/U=0.0$) (d) - near QM-SF ($zt/U=0.101$) (e) - SF ($zt/U=0.12$) (f) at $\mu/U_{0}=1.0$ for $\lambda/U=0.2$. In Fig.~\ref{fig:realspace}(d), one can easily understand that the majority of the lattice sites with $\tilde{\chi}>0.5$ remain in the MI phase, with one SF percolating cluster, which is essentially the signature of the QM phase. Subsequently, this QM phase gradually moves towards the SF phase, whence one or more SF percolating clusters appear with $\tilde{\chi}<0.5$ from a region near the QM-SF phase (e) to the SF phase (f).

Fig.~\ref{fig:realspace} (bottom row) includes the phase evolution from MI ($zt/U=0.001$) (g) - BG ($zt/U=0.017$) (h) - SF ($zt/U=0.022$) (i) at $\mu/U_{0}=1.0$ for $\Delta/U=0.2$. In the case of a random potential, there is no indication of the occurrence of the QM phase, since both in the MI and BG phases, majority of the lattice sites are in the insulating phase. There is no formation of a SF percolating cluster. This aids us to conclude that the quasiperiodic potential is solely responsible for the appearance of the QM phase. 

In order to understand the QM and BG phases more rigorously, and to get an idea of the first percolating cluster, we shall compute the mean cluster size which is defined as \cite{Stauffer},
\begin{equation}
M_{cs}(\chi)=\frac{\sum_{p}^{\infty} p^{2} s_{p}(\chi)}{\sum_{p}^{\infty} p s_{p}(\chi)}
\end{equation}
where $ps_{p} (\chi)$ represents the total number of the occupied sites corresponding to the $p$-th cluster and the percolating clusters are not included in the sum. Here, we shall employ the well known Hoshen-Kopelman algorithm that is mostly used in percolation problems. It works under the union-find principle to compute the the SF cluster formation \cite{PhysRevB.14.3438}.

\begin{figure}[t!]
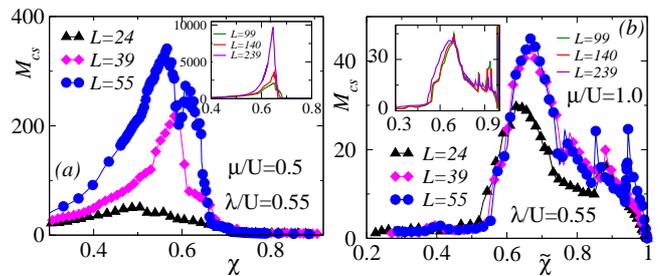

\centerline{\hfill
\includegraphics[width=0.24\textwidth]{Fig4d.eps}
\hfill
\hfill
\includegraphics[width=0.245\textwidth]{Fig4e.eps}
\hfill
}
\caption{The Mean cluster size, $M_{cs}$ as a function of $\chi (\tilde{\chi})$ in the QM-SF phase at $\lambda/U=0.55$ for $\mu/U=0.5$ in (a) and $\mu/U=1.0$ in (b).}
\label{fig:meanclustersizeqm}
\end{figure}

The variation of $M_{cs}$ corresponding to different lattice sizes is shown for the BG-SF in Fig.~\ref{fig:meanclustersizebg} and the QM-SF phase transitions in Fig.~\ref{fig:meanclustersizeqm} corresponding $\lambda/U=\Delta/U=0.55$. 
In Fig.~\ref{fig:meanclustersizebg}(a)-(c), the mean cluster size for BG-SF phase shows a similar trend for both types of disorder. At first, $M_{cs}$ increases as a function of $\chi$ and attains a maximum at a critical value, $\chi_{c}(L)$. Beyond this, $\chi_{c}(L)$, it starts to decrease hinting at the formation of the first percolating cluster across the lattice. This behaviour seems to be analogous to that of the 2D random percolation problem, where $M_{cs}$ attains a peak value at a critical threshold $\chi_{c}(L \to \infty)=0.592$. Thus we can infer that the values of $\chi_{c}(L)$ are $0.546~,0.621,$ and $0.674$, and thus it moves away from the critical point $\chi_{c}(L \to \infty)$ [Fig.~\ref{fig:meanclustersizebg}(a)] with increasing system sizes. Here the corresponding system sizes are chosen as $L=24,39,$ and $55$ respectively for the QP disorder. For the random disorder, the values for $\chi_{c}(L)$ are $0.472,~0.494,$ and $0.537$ corresponding to three system sizes, namely, $L=32,~64,$ and $128$ respectively, and they move towards $\chi_{c}(L \to \infty)$ [Fig.~\ref{fig:meanclustersizebg}(c)].  This clearly indicates that system experiences a finite size effect which needs to be dealt with. We discuss this in the next subsection.

%\begin{figure}[h!]
%\centerline{\hfill
%\includegraphics[width=0.35\textwidth]{4d.eps}
%\hfill}
%\centerline{\hfill
%\includegraphics[width=0.35\textwidth]{4e.eps}
%\hfill
%}
%\caption{The Mean cluster size, $M_{cs}$ as a function of $\chi (\tilde{\chi})$ in the QM-SF phase at $\lambda/U=0.55$ for $\mu/U=0.5$ in (a) and $\mu/U=1.0$ in (b).}
%\label{fig:meanclustersizeqm}
%\end{figure}

We have shown the mean cluster size for QM-SF phase transition corresponding to $\mu/U=0.5$ in Fig.~\ref{fig:meanclustersizeqm}(a) and $\mu/U=1.0$ in Fig.~\ref{fig:meanclustersizeqm}(b) in presence of QP potential of strength, $\lambda/U=0.55$. For $\mu/U=0.5$ [Fig.~\ref{fig:meanclustersizeqm}(a)], $M_{cs}$ is nearly symmetric with respect to variation in $\chi$, and the overall behaviour remains similar to that of the BG-SF phase as shown in Fig.~\ref{fig:meanclustersizebg}. It attains a maximum value at $\chi_{c}(L)= 0.513,~0.583,~$ and ~$0.567$ for the system sizes stated above, which is again a value close to $\chi_{c}(L \to \infty)$. For $\mu/U=1.0$ [Fig.~\ref{fig:meanclustersizeqm}(b)], $M_{cs}$ is asymmetric with respect to $\tilde{\chi}$ and the overall behaviour appears to be different from the BG-SF phase [Fig.~\ref{fig:meanclustersizebg}]. It shows multiple peaks in the region $\tilde{\chi}>0.5$ after attaining a maximum value at $\tilde{\chi}_{c}(L)$ to be $0.634,~0.671,$ and $0.673$ which is smaller than $\chi_{c}(L \to \infty)$. However, the magnitude of $M_{cs}$ in Fig.~\ref{fig:meanclustersizeqm}(b) is much less as compared to the other phases, possibly indicating that in the QM phase, the formation of a SF cluster is negligible. This happens since majority of the lattice sites now belong to the MI phase, along with a small patch of the SF percolating cluster. This is contrast with the BG phase having a large number of the SF clusters being surrounded by the MI clusters. 

However, we have observed that $M_{cs}$ becomes maximum at a single $\chi_{c}(L)$ for large system sizes for both the BG-SF [Fig.~\ref{fig:meanclustersizebg}(b)] and the QM-SF [Fig.~\ref{fig:meanclustersizeqm}(inset)] phase transitions. We thus can infer that there is no explicit finite size dependence of the system in the limit $L\ge 99$ for the QP disorder. 
\begin{figure}[b!]
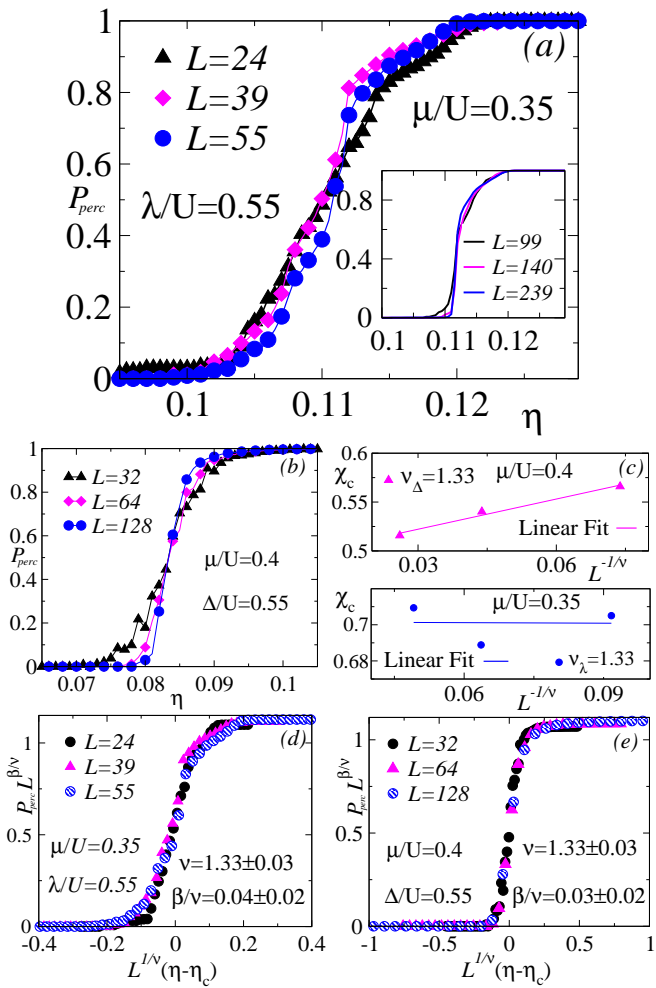

\centerline{\hfill
\includegraphics[width=0.40\textwidth]{Fig5a.eps}
\hfill
}
\centerline{
\hfill
\includegraphics[width=0.24\textwidth]{Fig5b.eps}
\hfill
\includegraphics[width=0.245\textwidth]{Fig5c.eps}
\hfill
\hfill}
\centerline{
\hfill
\includegraphics[width=0.245\textwidth]{Fig5e.eps}
\hfill
\hfill
\includegraphics[width=0.245\textwidth]{Fig5f.eps}
\hfill
}
\caption{The percolation probability, $P_{perc}$ 
  as a function of $\eta=zt/U$ in the BG-SF phase for different lattice size corresponding to QP in (a) and RP in (b). $\chi_{c}(L)$ vs $L^{-1/\nu}$ plot for QP and random disorder in (c). Finite-size scaling for QP disorder is shown in (d) and for random disorder in (e).}
\label{fig:finite_sizebg}
\end{figure}

\subsection{Finite-size scaling and critical exponents}
Endowed with the understanding of the mean cluster size and how to deal with the finite size effects, it is therefore necessary to look into the extent of the percolating cluster in real space at the onset of the BG-SF or QM-SF phases. For that purpose, we define another quantity, $P_{perc}$ via \cite{Stauffer,ROY2018969,PhysRevE.95.010101},
\begin{equation}
P_{perc}=\frac{{\rm{Sites~in~a~spanning~cluster}}}{{\rm{Total~number~of~occupied~sites}}}.
\end{equation}

\begin{figure}[t!]
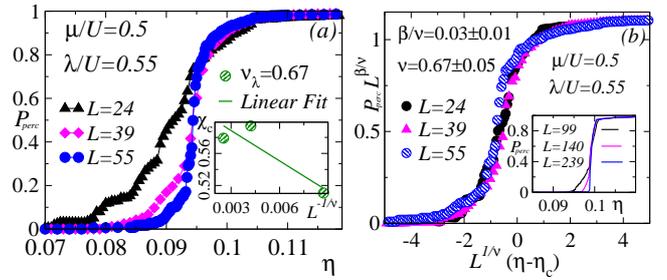

\centerline{
\hfill
\includegraphics[width=0.255\textwidth]{Fig6a.eps}
\hfill
\hfill
\includegraphics[width=0.23\textwidth]{Fig6b.eps}
\hfill
}
\caption{The Percolation probability, $P_{perc}$ 
  as a function of $\eta=zt/U$ in the QM-SF phase for different lattice sizes corresponding to (a) and finite-size scaling in (b) for QP disorder.}
\label{fig:finite_sizeqm}
\end{figure}

It is expected that the $P_{perc}=0$ in both the MI and the BG phases, while $P_{perc} \neq 0$ for the SF and the QM phases. To explore the finite size effects, the variation of $P_{perc}$ for the BG-SF phase is shown in Fig.~\ref{fig:finite_sizebg}, and the QM-SF phase in Fig.~\ref{fig:finite_sizeqm} corresponding to different system sizes, $L$ by considering the average over $sample=100$ in both the cases. 

For the BG-SF phase, we have shown $P_{perc}$ as a function of $zt/U$ (say, $\eta=zt/U$ ) for $\lambda/U=0.55$ in Fig.~\ref{fig:finite_sizebg}(a) and $\Delta/U=0.55$ in Fig.~\ref{fig:finite_sizebg}(b). For the quasiperiodic case, all the three curves pass through a critical point, namely, $\eta_{c}\approx 0.1105$ for a smaller system size, namely, $L\le55$ [Fig.~\ref{fig:finite_sizebg}(a)]. For the random case, it is observed that, all the three curves intersect at a single point, yielding the critical hopping strength $\eta_{c}\approx 0.083$ [Fig.~\ref{fig:finite_sizebg}(b)]. Thus to get rid of the finite-size effect, it is required to search for a universal scaling function corresponding to both types of disorder.

Now, at the critical point, $\eta_{c}$, $P_{perc}$ follows a scaling behavior, which we define by \cite{Stauffer,ROY2018969,PhysRevE.95.010101},
\begin{equation}
P_{perc}(L,\eta)=L^{-\beta/\nu}\tilde{p}(\eta-\eta_{c})L^{1/\nu}
\label{perc_univ}
\end{equation}
where $\tilde{p}$ is the universal scaling function and $\beta$, $\nu$ are the critical exponents. By choosing a proper value of $\beta$, $\nu$, and with the values $\eta_{c}$ obtained earlier, the system shows a length invariance, where different curves collapse on to each other which validates the existence of a universal scaling function (see Eq.~\ref{perc_univ}). It can be shown that the average size $\langle S_{m} \rangle$ of the SF percolating cluster varies with system size, $L$ at the percolation threshold ($\chi_{c}$) via, $\langle S_{m} \rangle \approx L^{d_{f}}$ where $d_{f}$ is the fractal dimension of the SF percolating cluster. It is related to the critical exponents and the system dimension $d$ through $d_{f}=d-\beta/\nu$ \cite{Stauffer,ROY2018969,PhysRevE.95.010101}.       

In order to find a suitable value of the exponent $\nu$, we need to locate $\chi_{c}(L)$ at which a spanning cluster appears for the first time. This can be obtained through the mean value of the distribution defined by the derivative of $P_{perc}$ with respect to $\chi$, namely, $\sfrac{dP_{perc}}{d\chi(\eta)}$ via $\chi_{c}(L)=\int\nolimits_{0}^{1}\chi \frac{dP_{perc}}{d\chi}\propto L^{-1/\nu}$, where $\chi_{c}(L)$ can be identified as the value for which $\sfrac{dP_{perc}}{d\eta}$ corresponds to a maximum. 

We have plotted $\chi_{c}(L)$ against $L^{-1/\nu}$ and found that the best straight line fit is observed for $\nu_{\lambda}=1.33\pm 0.03$ for the QP disorder and $\nu_{\Delta}=1.33\pm 0.03$ for the random disorder [Fig.~\ref{fig:finite_sizebg} (c)]. The subscripts with the critical exponents $\nu$ is used to differentiate the two disorder potentials . These values are again close to the values corresponding to random disorder in 2D obtained via QMC studies \cite{Stauffer,Niederle_2013}. Finally, having obtained $\nu$ and $\eta_{c}$ values and choosing $\beta/\nu=0.04\pm 0.02$, a suitable data collapse is observed. Subsequently, the fractal dimension, $d_{f}$ appears to be same corresponding to the both types of the potential [Fig.~\ref{fig:finite_sizebg}(d-e)], namely, we get $d_{f}\approx 1.96$ for both the cases. Therefore, we can conclude that, the phase transitions from the BG to the SF phase belong to the same universality class corresponding to both the types of disorder.

For the QM-SF phase, the variation of $P_{perc}$ with $\eta=zt/U$ is shown in Fig.~\ref{fig:finite_sizeqm}(a) for $\mu/U=0.5$ in presence of the QP potential. Here, all the three curves cross at the critical hopping strength given by $\eta_{c}\approx 0.0963$ [Fig.~\ref{fig:finite_sizeqm}(a)]. $\chi_{c}(L)$ against $L^{-1/\nu}$ plot shows a best fit straight line for $\nu_{\lambda}=0.67\pm 0.05$ [Fig.~\ref{fig:finite_sizeqm}(a)(inset)]. With these value of $\nu$, and $\beta/\nu=0.03\pm 0.01$, we have achieved a perfect data collapse for the QM-SF phase [Fig.~\ref{fig:finite_sizeqm}(b)]. The value of such a critical exponent ($\nu$) calculated using our percolation based mean field approach agrees very well with that from the QMC results obtained in Ref.\cite{PhysRevA.91.031604} for the QM-SF phase transition. Moreover, the phase transition from the QM phase to the SF phase belongs to a different universality class as compared to the BG to the SF phase transition.

For large system sizes, all the curves fall onto each other, thereby justifying that no further scaling is necessary for the QP disorder for $L\ge 99$ corresponding to BG-SF [Fig.~\ref{fig:finite_sizebg}(a)(Inset)] and the QM-SF phase transitions [Fig.~\ref{fig:finite_sizeqm}(a)(Inset)].

\subsection{Phase Diagram}
\begin{figure}[h!]
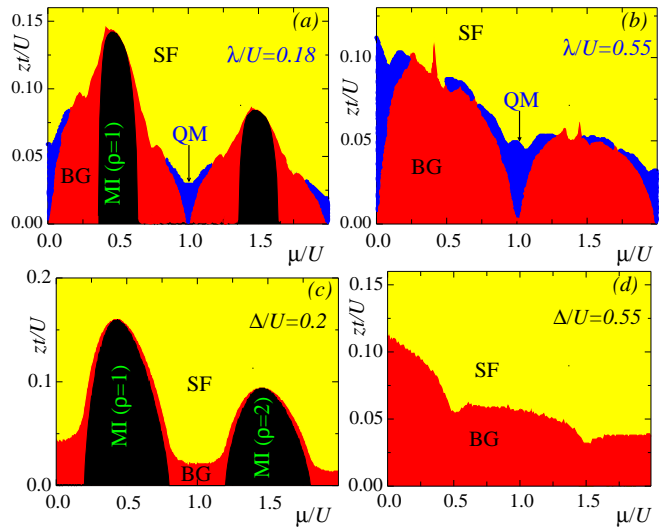

\centerline{\hfill
\includegraphics[width=0.245\textwidth]{Fig7a.eps}
\hfill
\hfill
\includegraphics[width=0.245\textwidth]{Fig7b.eps}
\hfill
}
\centerline{
\hfill
\includegraphics[width=0.24\textwidth]{Fig8a.eps}
\hfill
\hfill
\includegraphics[width=0.233\textwidth]{Fig8b.eps}
\hfill
}
\caption{Phase diagrams (top row) are shown for (a) $\lambda/U=0.18$ ,  and (b) $\lambda/U=0.55$  corresponding to QP disorder. While the phase diagrams (bottom row) are shown for (a) $\Delta/U=0.2$ and (b) $\Delta/U=0.55$  corresponding to random disorder. }
\label{fig:phasediagram}
\end{figure}

With the collective information obtained from $\tilde{\chi}$, $\chi$ and $P_{perc}$, we are now in the position to obtain the phase diagram corresponding to the random and the quasiperiodic potentials. The quantum phases can be characterized based on the SF percolating cluster which is summarized in Table \ref{table}. The resultant phase diagrams in the $\mu-zt$ plane are shown in Fig.~\ref{fig:phasediagram} where we have only scanned up to the second MI lobe for $sample=30$ realizations.
\begin{table}[htbp]
\begin{tabular}{cc|c|c|c|c|l}
\cline{3-6}
& & MI & QM & BG & SF \\
\hline
\multicolumn{1}{ |c  }{\multirow{1}{*}{$\chi~(\tilde{\chi})$} } 
 &  & $\tilde{\chi}=1$ & $\tilde{\chi}>0.5$ & $0<\chi<\chi_{c}$ &$\chi_{c}\le \chi \le 1$     \\ \cline{1-6}
\hline
\multicolumn{1}{ |c  }{\multirow{1}{*}{$P_{perc}$} } 
 &  & 0 & $\neq 0$ & 0 &$\neq 0$     \\ \cline{1-6}
\hline
\end{tabular}
\caption{Characterization of the quantum phases based on $\chi~(\tilde{\chi})$ and $P_{perc}$}
\label{table}
\end{table}

For $\lambda/U=0.18$, the phase diagram consists of all four phases, such as, the MI, BG, QM and the SF phases which are marked by different colours in Fig.~\ref{fig:phasediagram}(a). The presence of the QP disorder interrupts a direct MI-SF phase transition due to the appearance of the BG phase. The BG phase portrays a lobe structure similar to the MI lobe, and the chemical potential is now shifted by an amount $\pm 2\lambda/U$ on either side of the MI lobes. Apart form the BG phase, we have also found the signature of the QM phase, which forms like an envelope over the BG phase for small values of $\mu/U$, and is more pronounced at $\mu/U=1$, which is the degenerate point between the two consecutive MI lobes. 

With increasing strength of the QP disorder, the MI phase becomes more vulnerable, and the BG phase completely destroys the insulating phase at a critical value, $\lambda_{c}/U=0.25$. Further rise in the strength of the potential, for example to a value $\lambda/U=0.55$, the system consists of all the phases, except the MI phase, and the regions spanned by the BG and the QM phases dominate due to the effects of localization [Fig.~\ref{fig:phasediagram}(b)]. We have discussed it earlier that the QM phase is mostly insulating in nature having a small number of SF percolating cluster. Thus it behaves as a pseudo or a weak superfluid (wSF) phase and stabilizes with increasing QP strengths. All these phase diagrams are in qualitative agreement with the QMC \cite {PhysRevA.91.031604} and MFA results obtained in Ref.\cite{Johnstone_2021,johnstone2022barriers}, where the signature of the QM phase, namely, the wSF or the weak BG (wBG) is also observed, and the wSF phase seems to stabilize at larger values of $\lambda/U$. 

For $\Delta/U=0.2$, the disorder introduces the BG phase, and due to localization effects, the chemical potential is shifted up or down by an amount $\pm \Delta/2$ around the MI lobe [Fig.~\ref{fig:phasediagram}(c)]. On the contrary, the intervening BG region is less noticeable for small $\Delta/U$ as compared to the QP case, where the BG phase in the latter is distributed over a vast region around the MI phase. With increasing disorder strengths, the BG phase destroys the MI lobes at a critical value, $\Delta_{c}/U=0.5$, and the system only consists of the BG and SF phases beyond certain critical strength, $\Delta_{c}/U$ [Fig.~\ref{fig:phasediagram}(d)]. The phase diagrams are in qualitative agreement with the QMC and MFA results obtained in Ref.\cite{PhysRevA.99.053610,PhysRevB.85.020501,PhysRevA.91.043632,Niederle_2013}.

\section{Conclusion} 
In conclusion, we have studied a two-dimensional Bose-Hubbard model by considering two kinds of potential, namely, random and quasiperiodic (QP) potentials. While the random potential is fully uncorrelated, the quasiperiodic potential is deterministic in nature.
The inhomogeneity in the system leads to various quantum phases, such as Mott-insulator (MI), Bose-glass (BG), quasiperiodic mixed-phase (QM), and the superfluid (SF) phase.
A site decoupling mean field approach (MFA) in the context of percolation scenario is used to explore the quantum phases of the interacting ultracold atoms in optical lattices. The site inhomogeneity in the MFA is tackled via an $\it{indicator}$ for characterizing different quantum phases. Hence we have shifted our attention to the 2D random percolation problem to study the SF cluster distribution at the on-set of the BG/QM-SF phase transition. The appearance of the SF percolating cluster is studied using a well known Hoshen-Kopelman algorithm, and the mean cluster size attains a maximum value at a critical value close to the 2D random percolation threshold value. Further, the critical transition points and the exponents are calculated using finite-size scaling analysis corresponding to the BG/QM-SF phase transitions. We have found that our critical exponents are in good agreements with those obtained from the QMC results. Further, no explicit finite size dependence for large system sizes in case of the QP disorder case has been observed. Moreover, the BG to the SF phase transitions corresponding to random and QP disorder potentials belong to the same universality class. However, the QM to SF phase transition corresponding to the QP disorder belongs to different universality class.
Finally, the phase diagrams are obtained based on the critical value of the indicator, and the SF percolating cluster. The observation suggests that the BG phase appears in both cases, however, the QP disorder is more suitable than the random disorder in stabilizing a BG phase. In addition to that, the QP leads to the formation of a QM phase, which becomes robust with increasing QP potential strength. The QM phase is completely absent in the presence of a random potential.

\section{Acknowledgements}
SNN thanks B. Roy for useful discussions. SR acknowledges the Param-Ishan HPC at the IIT Guwahati for providing computational resources. SNN acknowledges the Param-Shakti and departmental HPC at the IIT Kharagpur for providing computational resources. SNN is supported by  IIT Kharagpur through IPDF grant No. IIT/ACD(PG S$\&$R)/PDF/offer/2021-22/PH.  
\bibliography{reference} \bibliographystyle{aip}

\begin{thebibliography}{10}
\bibitem{note1}
$M$ can be obtained via continued fraction approximation as discussed in
  Ref.\cite{Lang} .
\end{thebibliography}
\end{document}